\documentclass[3p,11pt,a4paper,review]{elsarticle}
\usepackage{lineno}
\modulolinenumbers[3]
\journal{}
\usepackage{amssymb,amsmath,latexsym,manfnt, amsfonts}
\usepackage{color}
\usepackage{float}
\usepackage{graphicx}
\usepackage{caption,subcaption}
\allowdisplaybreaks[4]
\usepackage[colorlinks,linkcolor=red,anchorcolor=blue,citecolor=green]{hyperref}

\usepackage{numcompress}\biboptions{authoryear}

\begin{document}

\begin{frontmatter}

\title{Non-Darcian flow or fractional derivative ?}

\author[]{H.W. Zhou\corref{mycorrespondingauthor}}
\cortext[mycorrespondingauthor]{Corresponding author. Tel: +86-10-62331286; Fax: +86-10-62331490.}
\ead{zhw@cumtb.edu.cn}

\author[]{S. Yang}
\author[]{R. Wang}
\author[]{J.C. Zhong}

\address{School of Mechanics and Civil Engineering,\\ China University of Mining and Technology, Beijing, 100083, P.R. China}

\begin{abstract}
Modeling of water and gas flow in low-permeability media is an important topic for a number of engineering such as exploitation of tight gas and disposal of high-level radioactive waste. It has been well documented in the literature that Darcy's linear law of water flux and the hydraulic gradient is not always adequate for low-permeability media. On the basis of the Swartzendruber model, as a non-Darcian model of flow in low-permeability media, a new nonlinear model of water flux and the hydraulic gradient is proposed using fractional derivative. The analytic solution of fractional derivative flow model is presented and all parameters of the fractional derivative flow model are determined by the Levenberg-Marquardt method on the basis of the experimental data of water flow in low-permeability media. It is indicated that the results estimated by the fractional derivative flow model proposed in the paper are in better agreement with the experimental data than the results estimated by the Swartzendruber equation. In addition, a sensitivity study is carried out, showing the effects of fractional derivative order and threshold hydraulic gradient on water flux.
\end{abstract}


\end{frontmatter}


\section{Introduction}
\noindent
Since Henry Darcy's remarkable modeling of linear relation between water flux and the hydraulic gradient, e.g., Darcian flow, in 1856, many researchers found the Darcy's law is not good enough for description of water and gas flow in low-permeability media like clay and shale. Consequently, in past decades, extensive efforts have been devoted to modeling approaches of nonlinear relation between water flux and hydraulic gradient, called non-Darcian flow \citep{Liu2014,Liu2016}. Aiming at low-permeability media like clay, \cite{Miller1963} supposed a threshold gradient for water flow in clays to distinguish linear and nonlinear flow. They found that water flow rate is linearly related to hydraulic gradient at gradients above the threshold gradient, but no flow occurs below threshold gradient. \cite{Deng2007} suggested a new equation of nonlinear flow in saturated clays that can describe characteristics of flow curve of the nonlinear flow from low to high hydraulic gradients.

Generally speaking, non-Darcian flow can be described by nonlinear functions of water flux and hydraulic gradient such as exponential and power functions. \cite{Hansbo1960,Hansbo2001} proposed a power relationship between water flux and hydraulic gradient for non-Darcian flow in clay media. By analyzing data sets for water flow in clay soils, \cite{Swartzendruber1962} proposed an exponential function to validate Darcy's law, resulting in a nonlinear relation of water flux versus gradient. In order to capture the non-Darcian flow behavior, \cite{Liu2012} developed a new relationship between water flux and hydraulic gradient by generalizing the currently existing relationships. The new relationship is shown to be consistent with experimental observations for both saturated and unsaturated conditions.

To validate Darcy's law and develop non-Darcian models seem to be an endless challenge.  It therefore leads to a new channel.  According to Darcy's law, water flux is directly proportional to the hydraulic gradient, i.e., the first order (as an integral number) derivative of water head with respect to the flow distance. In other words, Darcian flow can be described by an integer derivative of water head. Non-Darcian flow in porous media as a nonlinear phenomenon requires a new mathematical approach. In this case, non-Darcian flow could be characterized by a fractional derivative.

The fractional calculus, referred to as calculus of integrals and derivatives of any arbitrary real or complex order, is a 300 years old mathematical discipline. Its original conception is believed to have stemmed from a question raised in the year of 1695 by Marquis de L'H\^{o}pital (1661-1704) to Gottfried Wilhelm Leibnitz (1646-1716), the founder of Calculus. In the past few decades, the fractional calculus has gained remarkable popularity and importance because of its demonstration applications in numerous seemingly diverse and widespread fields of science and engineering \citep{Herrmann2011, Ortigueira2011} such as applications of fractional calculus to time-dependent behavior of rocks \citep{Zhou2011,Zhou2013} and composites \citep{Zhou2017}, fluid mechanics \citep{Kulish2002}, and solid mechanics \citep{Carpinteri2002,Carpinteri2004,Rossikhin2010}.

Moreover, some researchers devoted themselves to a nonlinear modeling approach of fractional derivative to non-Darcian flow. \cite{He1998} proposed a new model for seepage flow in porous media to modify the Darcy's law with fractional derivatives. \cite{Tian2006} researched the flow characteristics of fluids through a fractal reservoir with the fractional order derivative. By regarding the water flow as a function of a fractional derivative of the piezometric head, \cite{Cloot2006} generalized the classical Darcy's law to derive a new equation of groundwater flow. \cite{Chen2013} developed a new variable-order fractional diffusion equation to describe the diffusion process of chloride ions in the reinforced concrete structure.  \cite{Babak2014} presented a unified fractional differential approach to modeling flows of slightly compressible fluids through naturally fractured media. Recently, \cite{Wang2015} applied Caputo fractional constitutive equation to describe the transient electro-osmotic flow of a generalized Maxwell fluid in a cylindrical capillary.

As described by \cite{Cloot2006}, the underlying basic assumption of the fractional derivative modeling approach to transport in porous media is that the fluid flow at a given point of the porous media is governed not only by the properties of the piezometric field at the specific position but also depends on the global spatial distribution of that field in soil matrix. As a consequence, time or space fractional derivatives are extensively used in models of solute transport in porous media in order to take into account the memory effect or nonlocal properties induced by the interactions of fluid particles with pores of the porous media. Nevertheless, time or space fractional derivative models usually need to make dimensionless for convenience. Therefore, a different perspective to address this problem will be shown herein to interleave with the fractional calculus.

This paper represents an attempt to describe non-Darcian flow mathematically. The Swartzendruber equation as a non-Darcian flow model is generalized to describe the relation between water flux and hydraulic gradient using fractional derivative, resulting in a new model called the fractional derivative flow model. The analytic solution of fractional derivative flow model is presented and all parameters of the fractional derivative flow model are determined on the basis of the experimental data of water flow in low-permeability media. The results estimated by the fractional derivative flow model proposed in the paper are in better agreement with the experimental data than the results estimated by the Swartzendruber model. It indicates that our perspective of fractional derivative modeling approach is acceptable for non-Darcian flow in porous media.
\section{Fractional derivative approach to non-Darcian flow}
\subsection{Definition of the Caputo derivative}
The some definitions of fractional derivatives are popular in mathematics like Grunwald-Letnikov, Riemann-Liouville, and Caputo derivative \citep{Podlubny1999}. Among them, Caputo derivative is widely used in physics and mechanics because of its advantages in solving fractional differential equations with initial conditions. For a given function $f(x)$ Caputo derivative is defined by
\begin{equation}\label{Eq.(1)}
  \frac{{{d^\gamma }f(x)}}{{d{x^\gamma }}} = \frac{1}{{\Gamma (n - \gamma )}}\int_0^x {\frac{{{f^{(n)}}(t)}}{{{{(x - t)}^{\gamma  - n + 1}}}}dt},
\end{equation}
where $\gamma>0$, $n$ is the least integer greater than $\gamma$, and $\Gamma(\cdot)$ is the Gamma function, i.e., $\Gamma (\gamma ) = \int_0^\infty  {{t^{\gamma  - 1}}{e^{ - t}}dt}$. In particular, for $\gamma=0$, $\frac{{{d^\gamma }}}{{d{x^\gamma }}}$ denotes the identity operator.
\subsection{ Darcian flow}
Considering one-dimensional steady-state flow, suppose a fluid flows along a straight line, say, $x$-direction, the flux is related to hydraulic gradient by a well-known Darcy's law given by
\begin{equation}\label{Eq.(2)}
q = \frac{K}{\mu }\frac{{dp}}{{dx}},
\end{equation}
where $q$ is the bulk velocity of fluid $(\mathrm{m}/\mathrm{s})$, or fluid flux, $K$ is permeability $(\mathrm{m}^2)$, $\mu$ is dynamic viscosity $(\mathrm{N}\cdot\mathrm{s}/\mathrm{m}^2)$, $p$ is fluid pressure $(\mathrm{N}/\mathrm{m}^2)$.

Eq.(\ref{Eq.(2)}) can be usually described by
\begin{equation}\label{Eq.(3)}
 q = ki,
\end{equation}
where $k = \frac{K}{\mu }\rho$ is hydraulic conductivity $(\mathrm{m}/\mathrm{s})$, $\rho$ is density of fluid $(\mathrm{N}/\mathrm{m}^3)$, and $i$ is hydraulic gradient.
\subsection{Non-Darcian flow}
Non-Darcian flow, generally, can be described by nonlinear functions of water flux and hydraulic gradient such as power and exponential functions.

(1) Power function

Darcy's law leads to a linear relation between $q $ and $i$ as shown in Eq.(\ref{Eq.(3)}). By producing first order derivative to both sides of Eq.(\ref{Eq.(3)}), we then have a differential equation like
\begin{equation}\label{Eq.(4)}
dq = kdi  \quad or \quad \frac{{dq}}{{di}} = k.
\end{equation}

It is shown that Darcian flow can be described by an integer derivative of flux with respect to the hydraulic gradient $i$, as a dimensionless variable.  A similar model is the Newtonian dashpot for description of a linear relationship between the viscous stress and the rate of strain. The Newtonian dashpot was developed to the Abel dashpot by invoking the fractional derivative \citep{Scott-Blair1944,Kiryakova1999,Zhou2011,Zhou2013}. In an analogous way, we suppose non-Darcian flow can be described by a fractional derivative of flux, which leads to a dimensionless form, i.e.,
\begin{equation}\label{Eq.(5)}
 \frac{{{d^\gamma }q}}{{d{i^\gamma }}} = k,\quad \gamma>0,
\end{equation}
where $\frac{{{d^\gamma }}}{{d{i^\gamma }}}$ is the Caputo fractional derivative operator.

Applying the bilateral Laplace transform $(LT)$ to Eq.(\ref{Eq.(5)}) gives
\begin{equation}\label{Eq.(6)}
LT\left[\frac{{d^\gamma }q}{d{i^\gamma }}\right] = {s^\gamma } \tilde q(s) - \sum\limits_{j = 0}^{n - 1} {{s^{\gamma - j - 1}}\frac{{{d^j}q(0)}}{{d{i^j}}}}  = \frac{k}{s}.
\end{equation}
Let $q(0) = 0$, we have:
\begin{equation}\label{Eq.(7)}
\tilde q(s){\rm{ = }}\frac{k}{{{s^{\gamma  + 1}}}}.
\end{equation}

Applying the inverse Laplace transform to Eq.(\ref{Eq.(7)}), i.e.,
\begin{equation}\label{Eq.(8)}
  L{T^{ - 1}}[\tilde q(s)]{\rm{ = }}L{T^{ - 1}}\left[\frac{k}{{{s^{\gamma  + 1}}}}\right] = \frac{k}{{\Gamma (1 + \gamma )}}{i^\gamma },
\end{equation}we have:
\begin{equation}\label{Eq.(9)}
q = k\frac{{{i^\gamma }}}{{\Gamma (1 + \gamma )}}.
\end{equation}
In this case, we get a power function of water flux $q$ and hydraulic gradient $i$ in Eq.(\ref{Eq.(9)}), showing a similar form of nonlinear equation supposed by \cite{Hansbo1960, Hansbo2001}.

(2) Exponential function: Fractional Swartzendruber equation

\cite{Swartzendruber1962} proposed an exponential relation between water flux and hydraulic gradient to modify Darcy's law, i.e.,
\begin{equation}\label{Eq.(10)}
\frac{{dq}}{{di}} = k(1 - {e^{ - \frac{i}{I}}}).
\end{equation}

Integrating both sides of Eq.(\ref{Eq.(10)}) and considering $q(0)=0$, we have:
\begin{equation}\label{Eq.(11)}
 q = k[i - I(1 - {e^{ - \frac{i}{I}}})]
\end{equation}
where $I$ is the threshold gradient and actually refers to the intersection of the linear part in plot of the hydraulic gradient and the water flux.

Replacing integer derivative with fractional derivative in Eq.(\ref{Eq.(10)}), we have the fractional derivative Swartzendruber equation, i.e.,
\begin{equation}\label{Eq.(12)}
 \frac{{{d^\gamma }q}}{{d{i^\gamma }}} = k(1 - {e^{ - \frac{i}{I}}}),\quad 0\leq\gamma\leq1.
\end{equation}

Application of the Laplace transform $(LT)$ to Eq.(\ref{Eq.(12)}) leads to:
\begin{equation}\label{Eq.(13)}
{s^\gamma } \tilde q(s) = k\left( {\frac{1}{s} - \frac{1}{{s + {1 \mathord{\left/
 {\vphantom {1 I}} \right.
 \kern-\nulldelimiterspace} I}}}} \right),
\end{equation}then we have:
\begin{equation}\label{Eq.(14)}
  \tilde q(s) = \frac{{k{s^{ - \gamma  - 1}}}}{{1 + Is}}.
\end{equation}

Applying the inverse Laplace transform to Eq.(\ref{Eq.(14)}) like
\begin{equation}\label{Eq.(15)}
 L{T^{ - 1}}[\tilde q(s)]{\rm{ = }}L{T^{ - 1}}\left[\frac{{k{s^{ - \gamma  - 1}}}}{{1 + Is}}\right] = \frac{k}{I}{i^{\gamma  + 1}}{E_{1,\gamma  + 2}}( - \frac{i}{I}),
\end{equation}we have:
\begin{equation}\label{Eq.(16)}
q = \frac{k}{I}{i^{\gamma  + 1}}{E_{1,\gamma  + 2}}( - \frac{i}{I}),
\end{equation}
where ${E_{1,\gamma  + 2}}(\cdot)$ refers to Mittag-Leffler function, i.e., ${E_{1,\gamma  + 2}}( - \frac{i}{I}) = \sum\limits_{k = 0}^\infty  {\frac{{{{( - \frac{i}{I})}^k}}}{{\Gamma (k + \gamma  + 2)}}}$  \citep{Mainardi2010}.

In the case of $\gamma=0$, using ${E_{1,2}}( - \frac{i}{I}) = \frac{{{e^{ - i/I}} - 1}}{{ - i/I}}$, Eq.(\ref{Eq.(16)}) can be rewritten as
\begin{equation}\label{Eq.(17)}
 q = \frac{k}{I}i{E_{1,2}}( - \frac{i}{I}) = k(1 - {e^{ - \frac{i}{I}}}),
\end{equation}
which appearance is the same as Eq.(\ref{Eq.(10)}) if differential order $\gamma=0$.

In addition, the case of $\gamma=1$ gives
\begin{equation}\label{Eq.(18)}
 {E_{1,3}}( - \frac{i}{I}) = \frac{{{e^{ - i/I}} - 1 + i/I}}{{{{(i/I)}^2}}}.
\end{equation}

Substituting Eq.(\ref{Eq.(18)}) into Eq.(\ref{Eq.(16)}), we have $q = k(i - I(1 - {e^{ - \frac{i}{I}}}))$ indicating that the Swartzendruber equation in Eq.(\ref{Eq.(11)}) is a special case of the fractional derivative flow model when the fractional derivative order $\gamma=1$.

\section{Parameter determination for fractional derivative non-Darcian model}
\subsection{Parameter determination}
The efficacy of the fractional derivative model is dependent on its ability to adequately fit experimental data. Using the experimental data of water flux with hydraulic gradient, the parameters $k, I, \gamma$ in Eq.(\ref{Eq.(16)}) can be determined by the Levenberg-Marquardt method, a nonlinear least-squares fitting (LSF) method (see \citealt{Zhou2011} for details).

In what follows, we now use the fractional derivative flow model to fit the experimental data \citep{Wang2016} by LSF analysis. \cite{Wang2016} developed an experimental study to investigate the non-Darcian behavior of water flow in soil-rock mixtures (SRM) with various rock block percentages. Their work presented the data set of water flux as a power function of hydraulic gradient. In addition, the relationship between threshold hydraulic gradient and rock block percentage was also considered. The exact value of threshold hydraulic gradient $I$ for SRM specimens are listed in \autoref{Table 1}. Consequently, only the two parameters $k$ and $\gamma$ are remain to be determined. The results of least-squares fit of the parameters in Eq.(\ref{Eq.(16)}) to the experimental data \citep{Wang2016} are listed in \autoref{Table 1}.
\begin{table}[!tb]
  \centering
  \caption{\small Determination of parameters for fractional derivative flow model based on SRM specimens}\label{Table 1}
  \scalebox{0.85}{
  \begin{tabular}[]{cccccccccc}
  \hline
  SRM 	&\multicolumn{4}{c}{Swartzendruber equation}	&	\multicolumn{5}{c}{Fractional derivative flow model}\\
  \cline{2-10}
specimens &  $k\times10^{-5}(\mathrm{m}/\mathrm{s})$ & $I$ &$\mathrm{R}^2$ &MSE	& $ k\times10^{-5}(\mathrm{m}/\mathrm{s})$	&  $I $	     &$\gamma$&	 $\mathrm{R}^2$ & MSE\\
\hline
SRM20-1	& 0.2039 &	141.00 &  0.9817	&0.2242	& 0.2039	& 141.00	 & 1	    &0.9817	 &0.2242\\
SRM30-1	& 0.1378 &  130.20 &  0.9898	&0.0531	& 0.1378	& 130.20	 & 1	    &0.9898	 &0.0531\\
SRM40-1	& 0.07383&	123.60 &  0.9869	&0.1358	& 0.1311	& 123.60	 &0.8501	&0.9908	 &0.0095\\
SRM50-1	& 0.1297 &  102.50 &  0.9326	&0.2322	& 0.1297	& 102.50	 & 1	    &0.9326	 &0.2322\\
SRM60-1	& 0.1175 &  85.33  &  0.9786	&0.0363	& 0.1954	& 85.33	     &0.8567	&0.9819	 &0.0307\\
SRM70-1	& 0.1402 &  73.59  &  0.9710	&0.0291	& 0.3223	& 73.59	     &0.7454	&0.9807	 &0.0193\\
  \hline
\end{tabular}}
\end{table}

\begin{figure}[!htbp]
  \centering
  \begin{subfigure}[t]{0.45\textwidth}
  \centering
  \includegraphics[width=1\textwidth]{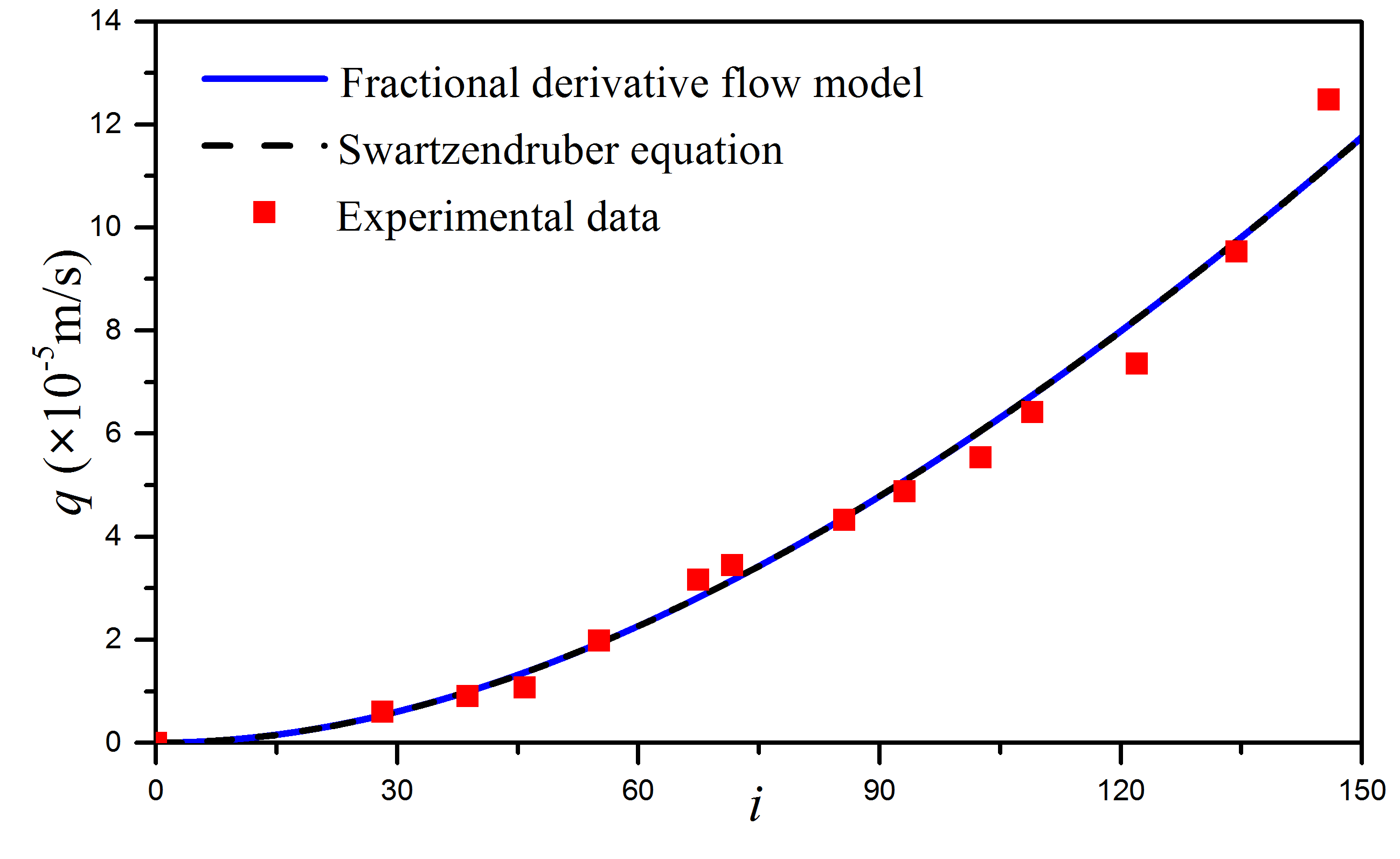}
  \subcaption*{\small(a)}
  \end{subfigure}
  \quad
  \begin{subfigure}[t]{0.45\textwidth}
  \centering
  \includegraphics[width=1\textwidth]{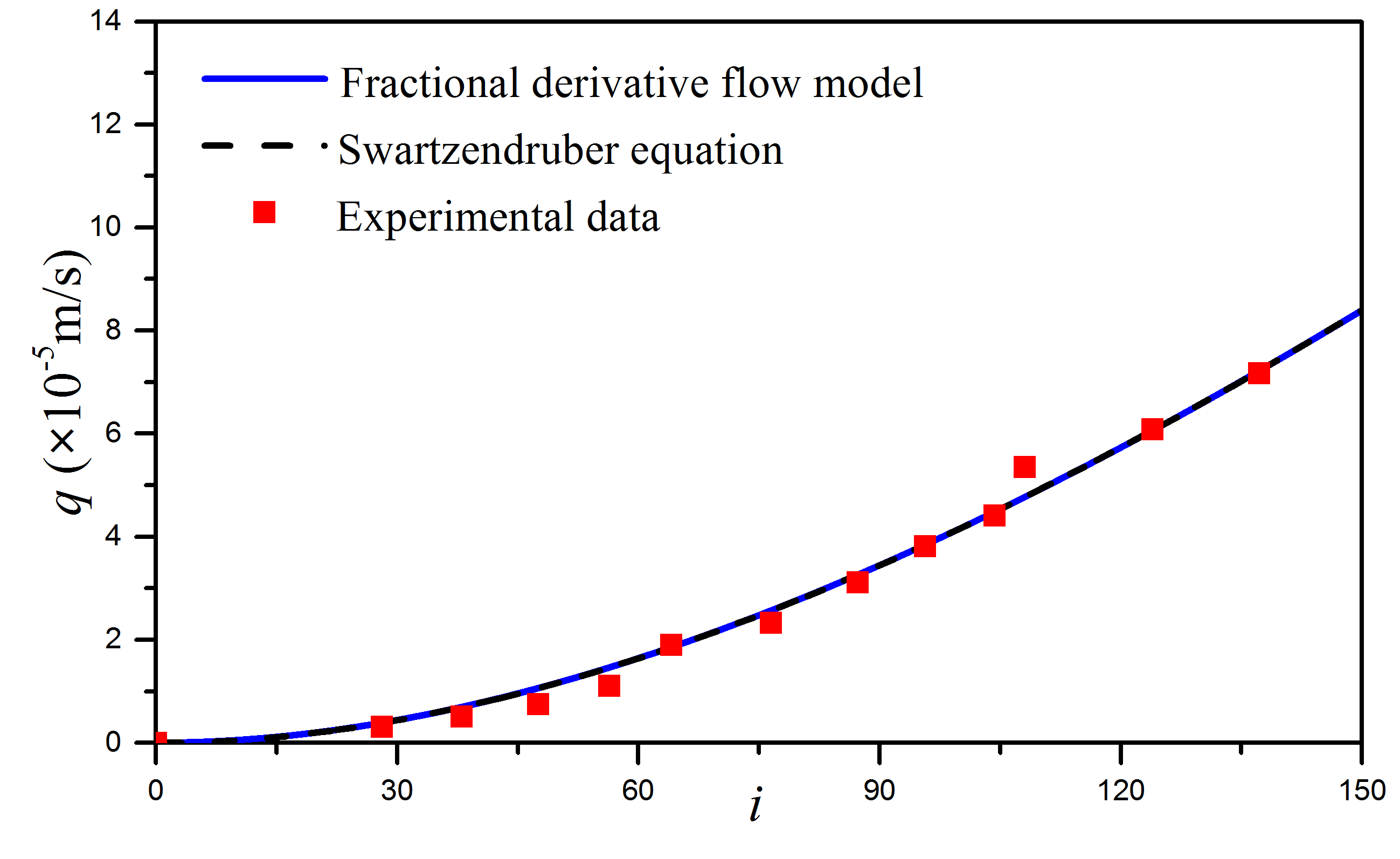}
  \subcaption*{\small(b)}
  \end{subfigure}
  \\
  \begin{subfigure}[t]{0.45\textwidth}
  \centering
  \includegraphics[width=1\textwidth]{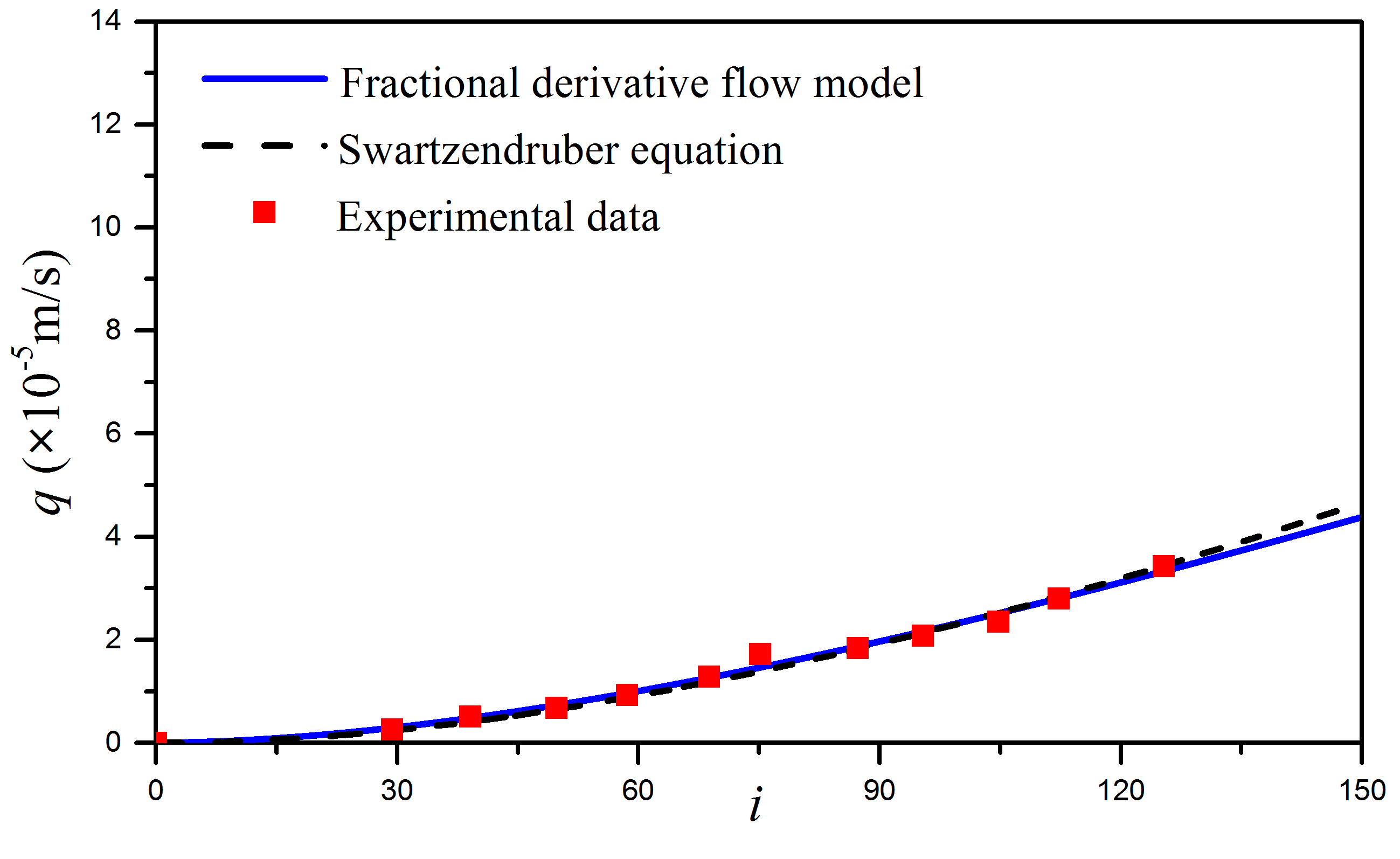}
  \subcaption*{\small(c)}
  \end{subfigure}
  \quad
  \begin{subfigure}[t]{0.45\textwidth}
  \centering
  \includegraphics[width=1\textwidth]{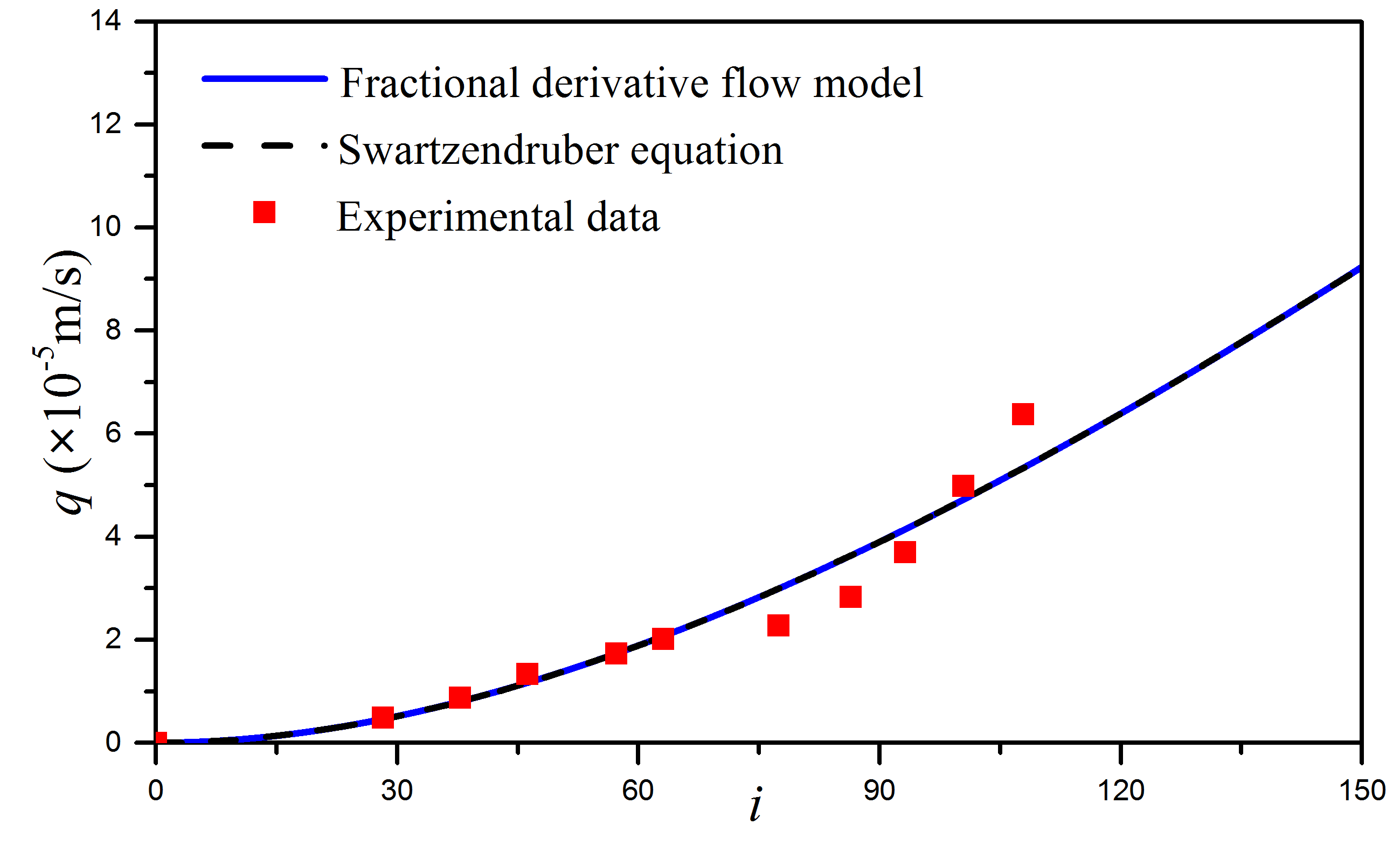}
  \subcaption*{\small(d)}
  \end{subfigure}
  \\
  \begin{subfigure}[t]{0.45\textwidth}
  \centering
  \includegraphics[width=1\textwidth]{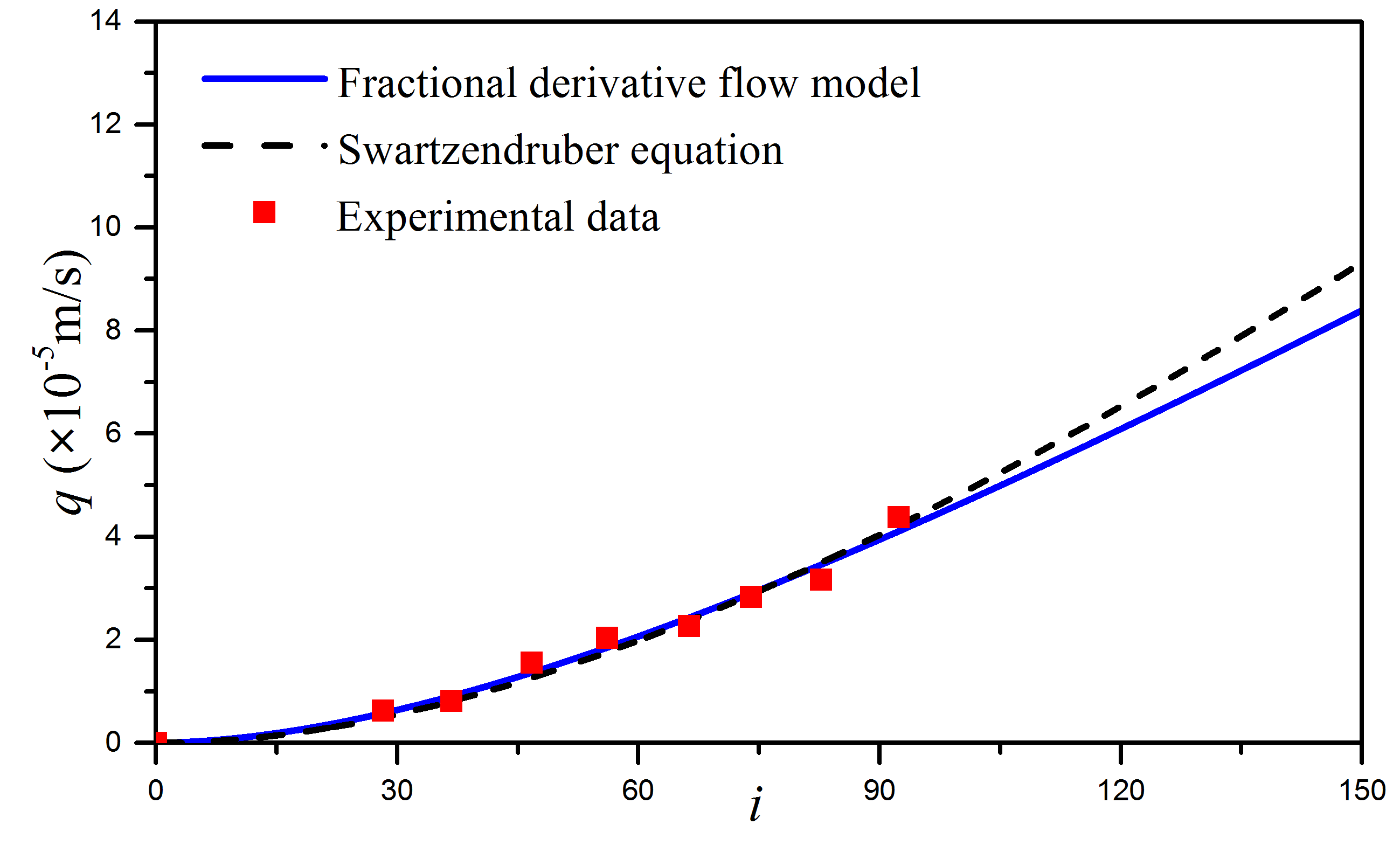}
  \subcaption*{\small(e)}
  \end{subfigure}
  \quad
  \begin{subfigure}[t]{0.45\textwidth}
  \centering
  \includegraphics[width=1\textwidth]{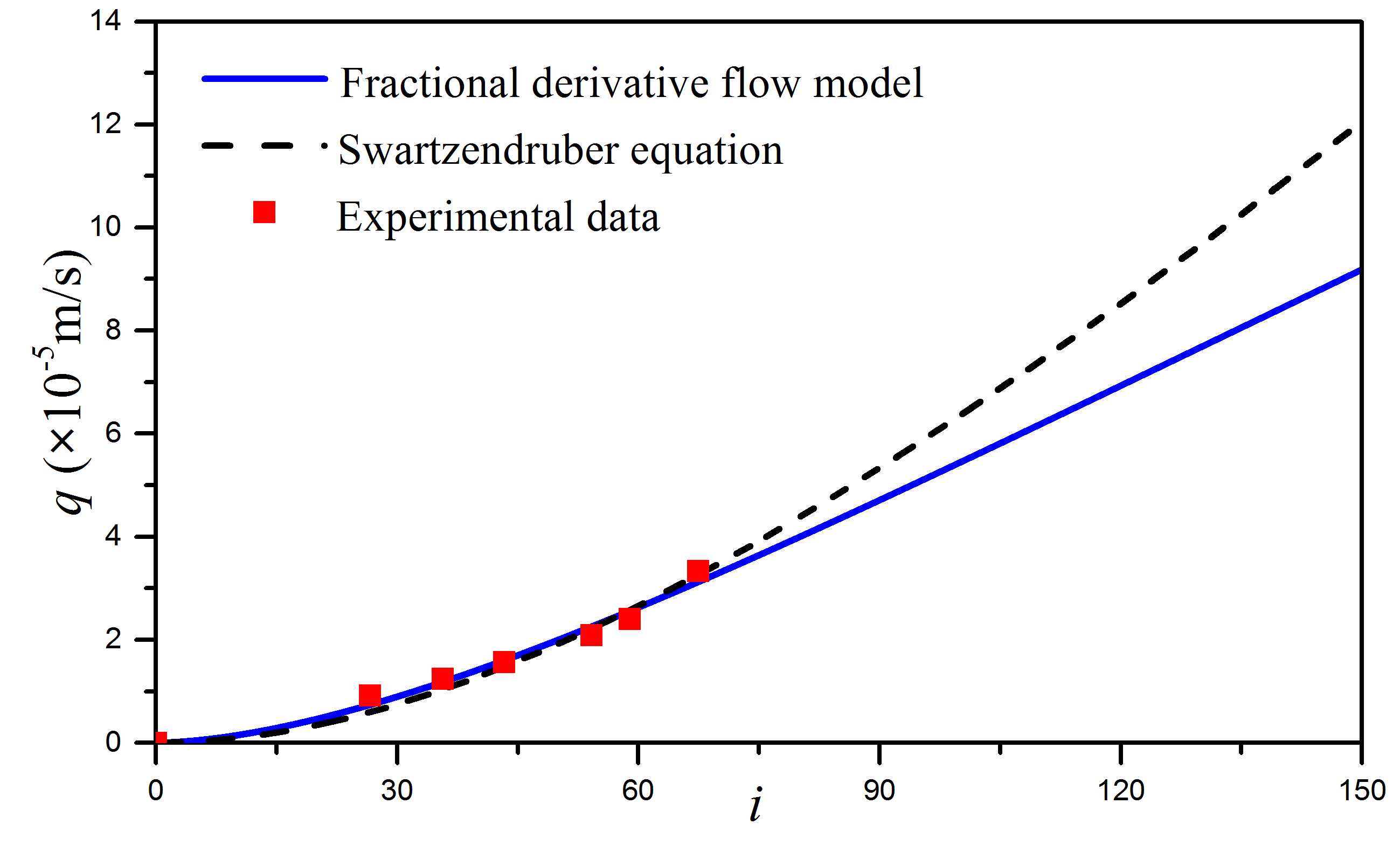}
  \subcaption*{\small(f)}
  \end{subfigure}
  \caption{\small Fitting curves theoretically given by fractional derivative flow model on the basis of experimental data for SRM specimens with different rock block percentages: 20\% (a), 30\% (b), 40\% (c), 50\% (d), 60\% (e) and 70\% (f) (see \citealt{Wang2016}).}
  \label{Fig1}
\end{figure}

The data as well as the fitting curves given by the fractional derivative flow model in Eq.(\ref{Eq.(16)}) is shown in \autoref{Fig1}. Making the fitting analysis to the same experimental data using the Swartzendruber equation in Eq.(\ref{Eq.(11)}), another set of parameters are given in \autoref{Table 1} as well. The least-squares analysis results in \autoref{Table 1} indicate that the fractional derivative flow model in Eq.(\ref{Eq.(16)}) is in better agreement with the experimental data than the Swartzendruber equation in Eq.(\ref{Eq.(11)}) with higher correlation coefficients ($\mathrm{R}^2$) and lower mean squared errors (MSE).

In addition, \autoref{Table 1} shows that an increase of rock block percentage in SRM specimens leads a decrease of the hydraulic conductivity $k$ to a minimum value at a rock block percentage of 40\%, and an increase if rock block percentage exceed 40\%. The similar behavior is also given in \cite{Wang2016}.

\begin{figure}[!ht]
  \centering
  \begin{subfigure}[t]{0.45\textwidth}
  \centering
  \includegraphics[width=1\textwidth]{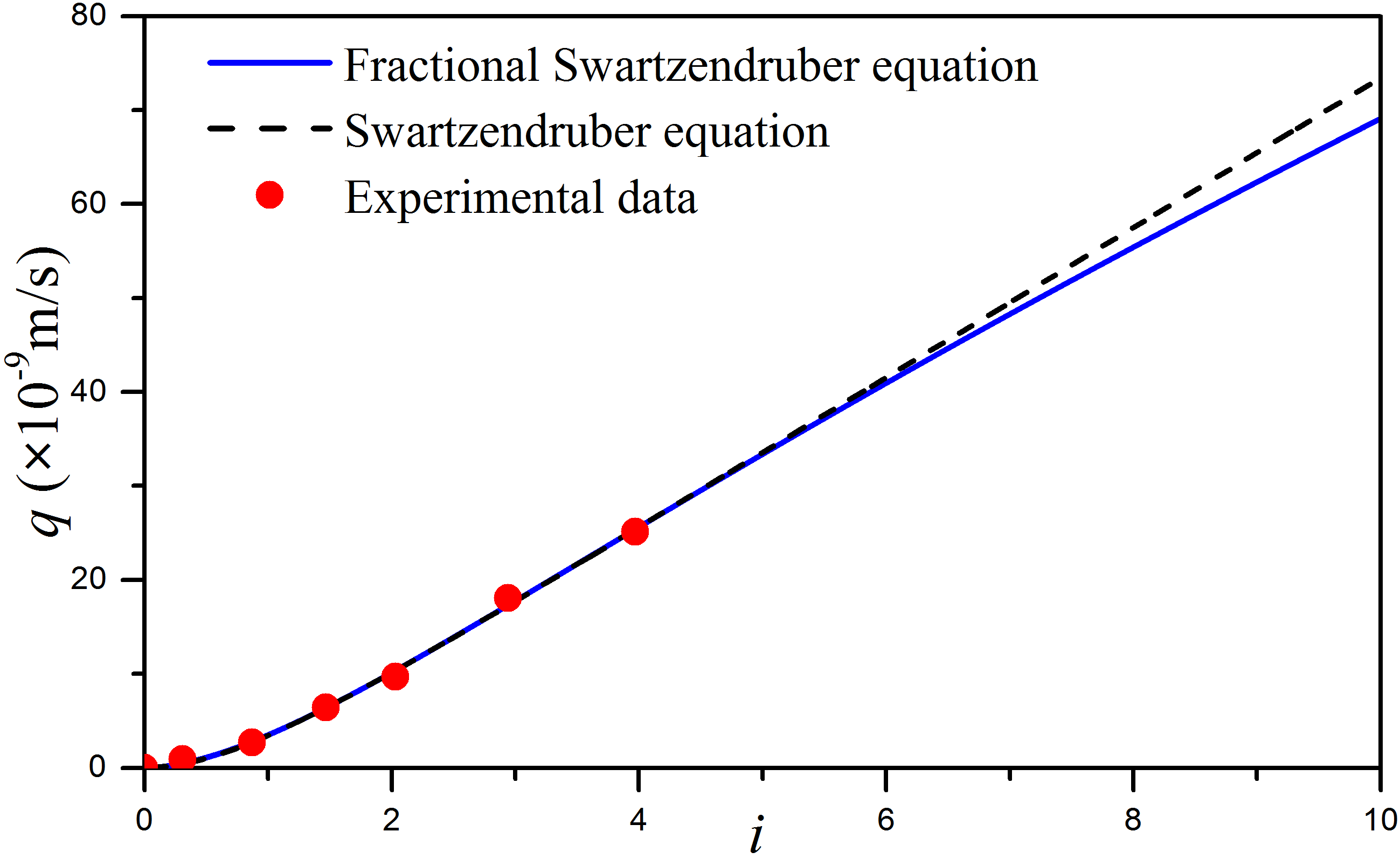}
  \subcaption*{\small(NO.64-3)}
  \end{subfigure}
  \quad
  \begin{subfigure}[t]{0.45\textwidth}
  \centering
  \includegraphics[width=1\textwidth]{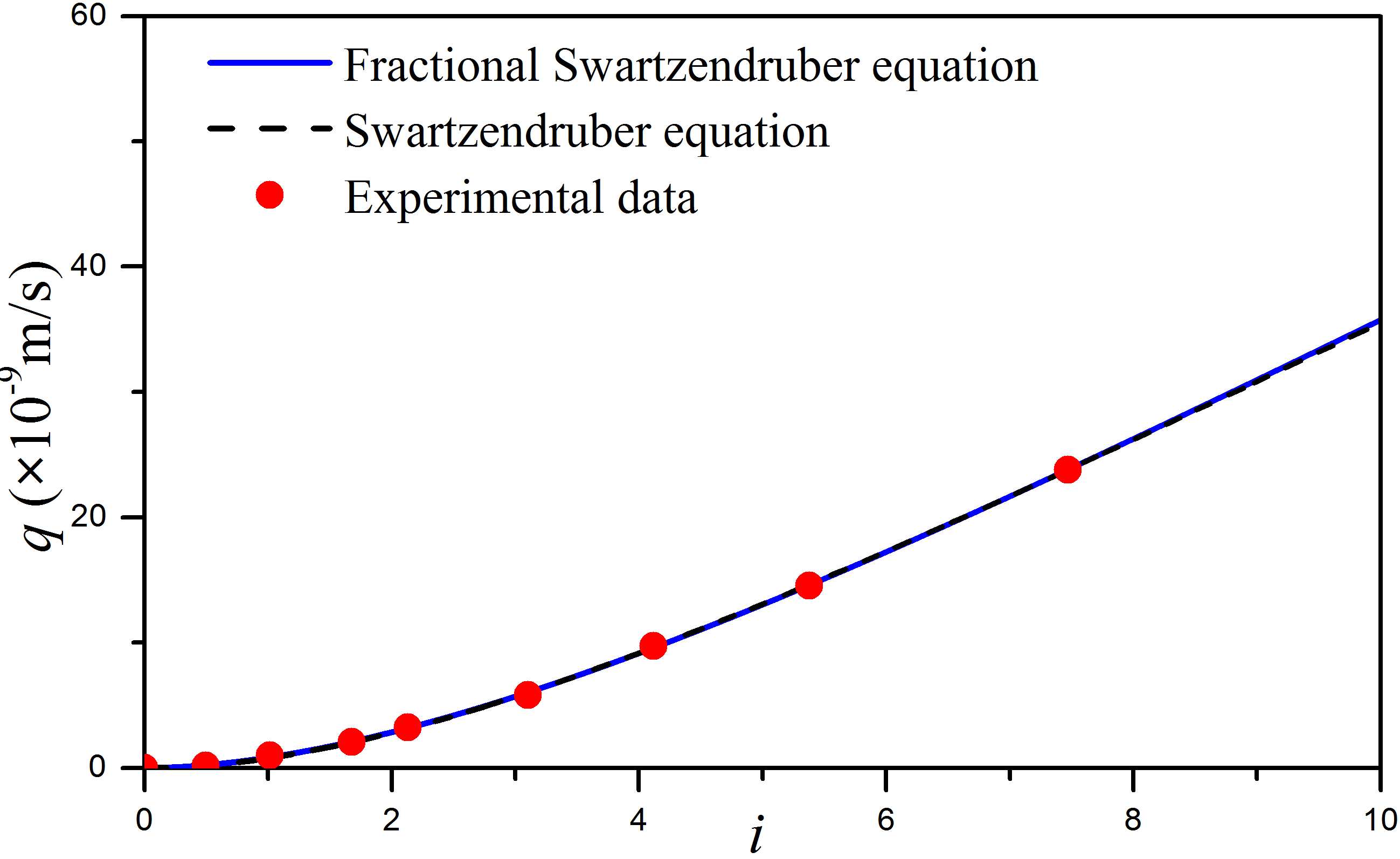}
  \subcaption*{\small(NO.64-4)}
  \end{subfigure}
  \caption{\small Fitting curves theoretically given by fractional derivative flow model on the basis of experimental data for saturated clays NO.64-3, NO.64-4 (see \citealt{Deng2007}).}\label{Fig2}
\end{figure}

Furthermore, using a more data set \citep{Deng2007}, the validity of our fractional derivative flow model was evaluated by LSF analysis. \cite{Deng2007} presented a nonlinear model of flow in saturated clays.
\begin{table}[!bht]
  \centering
  \caption{\small Determination of parameters for fractional derivative flow model based on saturated clays}\label{Table 2}
  \scalebox{0.85}{
  \begin{tabular}[]{cccccccccc}
  \hline
  Saturated 	&\multicolumn{4}{c}{Swartzendruber equation}	&	\multicolumn{5}{c}{Fractional derivative flow model}\\
  \cline{2-10}
 clays&  $k\times10^{-9}(\mathrm{m}/\mathrm{s})$ & $I$ &	$\mathrm{R}^2$ &MSE	& $ k\times10^{-9}(\mathrm{m}/\mathrm{s})$	&  $I $	     &$\gamma$&	 $\mathrm{R}^2$ & MSE\\
\hline
NO.64-3	& 7.973 &  0.7901 &  0.9973	&0.2050	& 11.52	& 1.565	 & 0.8094	  &0.9975	 &0.1501\\
NO.64-4	& 4.856 &  2.754  &  0.9998	&0.0116	& 7.153	& 4.408	 & 0.8814	  &0.9998	 &0.0100\\
\hline
\end{tabular}}
\end{table}
Comparisons of experimental data and fitting curves given by both Swartzendruber equation in Eq.(\ref{Eq.(11)}) and the fractional derivative flow model in Eq.(\ref{Eq.(16)}) are illustrated in \autoref{Fig2}. For better analysis, the parameters in Eq.(\ref{Eq.(11)}) and Eq.(\ref{Eq.(16)}) are determined and listed in \autoref{Table 2}. The results demonstrate that the proposed fractional derivative flow model is in better agreement with the experimental data than the Swartzendruber equation in Eq.(\ref{Eq.(11)}).

Moreover, since Swartzendruber equation in Eq.(\ref{Eq.(11)}) is a special case of the fractional derivative flow model when the fractional derivative order $\gamma=1$, our fitting results verified that the presented fractional derivative flow model is more flexible and accurate.

\subsection{Sensitivity analysis}
\noindent(1) Fractional derivative order

\begin{figure}[H]
  \centering
  \includegraphics[width=0.5\textwidth]{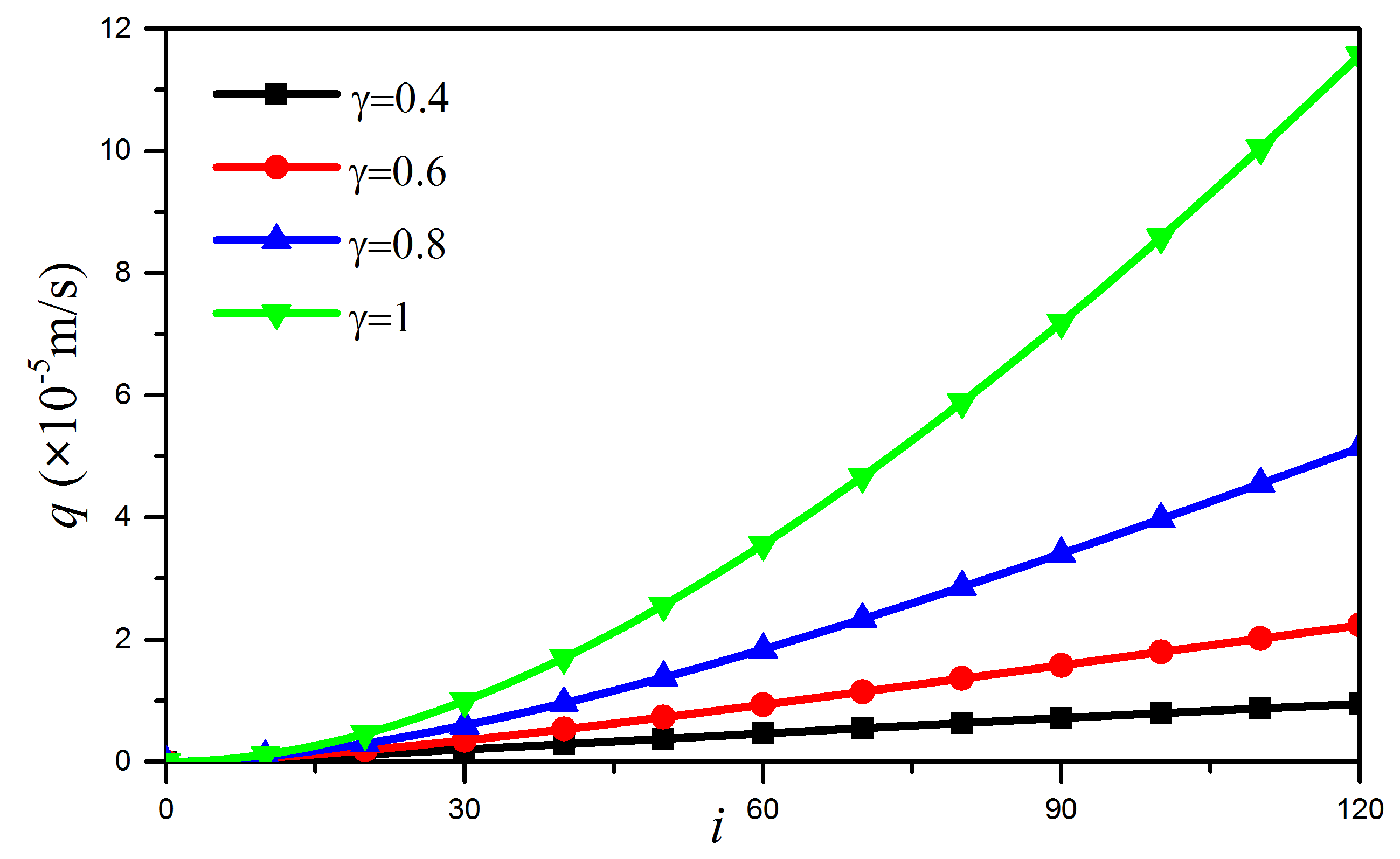}
  \caption{\small Sensitivity of the water flux to the fractional derivative order}\label{Fig3}
\end{figure}

Eq.(\ref{Eq.(16)}) shows that the relationship between water flux $q$ and hydraulic gradient $i$ depends on parameters $k, I, \gamma$. In order to get a better understanding of the effects of these parameters, sensitivity analyses have been carried out. The effect of fractional order $\gamma$ on the variation of water flux with hydraulic gradient is shown in \autoref{Fig3}. In which one parameter $\gamma$ takes three different values to show its effect on the $q-i$ curve under the condition of $k=2\times10^{-6}\mathrm{m}/\mathrm{s}, I=80$. It is shown that the higher the fractional derivative order, in general, the larger the fluid flux.

\noindent(2) Threshold hydraulic gradient

\begin{figure}[H]
  \centering
  \includegraphics[width=0.5\textwidth]{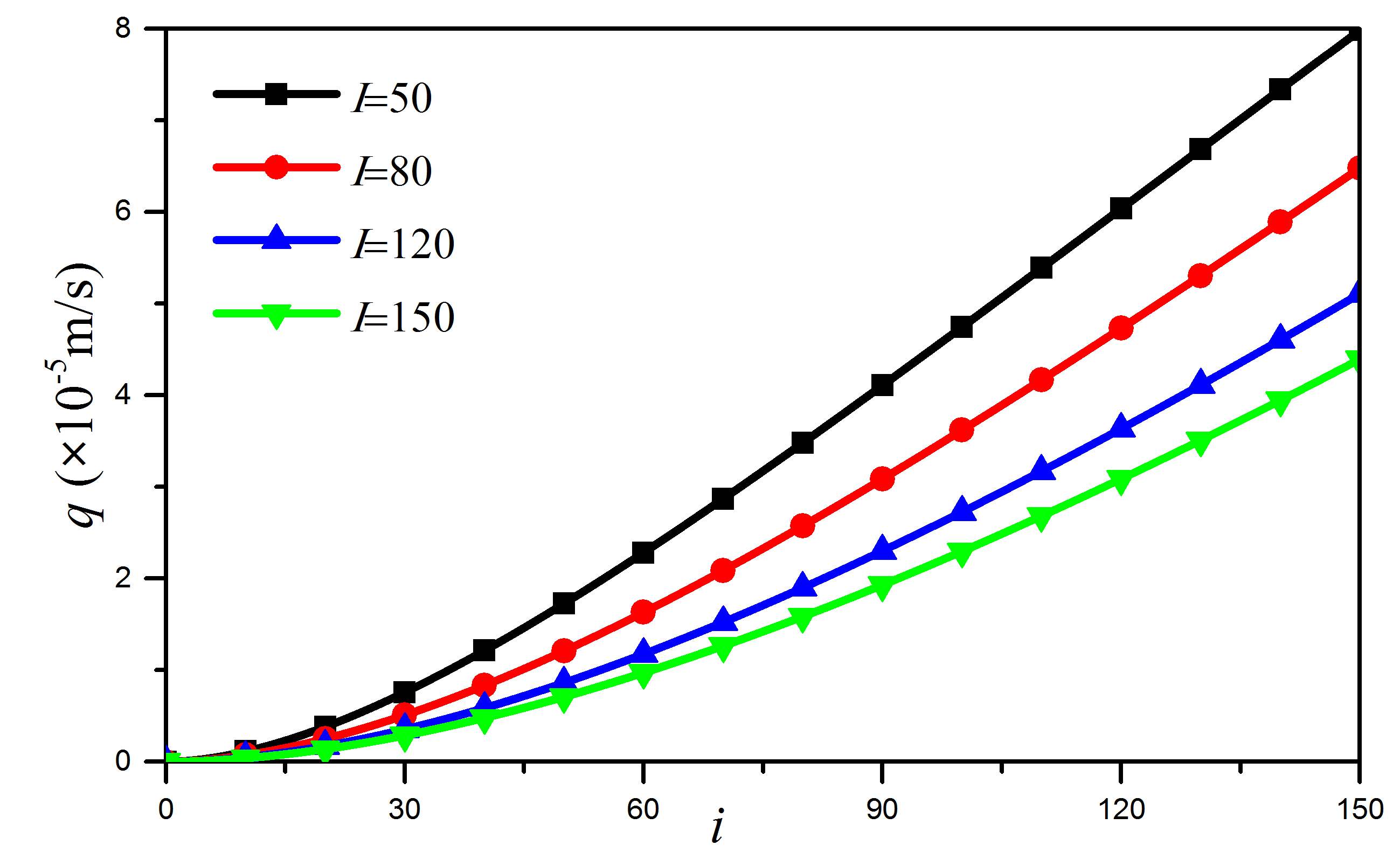}
  \caption{\small Sensitivity of the water flux to threshold hydraulic gradient}\label{Fig4}
\end{figure}

In Eq.(\ref{Eq.(16)}), let the threshold gradient $I$ changes and other parameters be constant, where $k=1.5\times10^{-6}\mathrm{m}/\mathrm{s}, \gamma=0.85$. A series of curves can be obtained as shown in \autoref{Fig4}, indicating that the higher threshold gradient, the smaller water flux.

\section{Conclusions}
The object of the present work is to develop fractional order equations for describing non-Darcian behavior between water flux and hydraulic gradient. Based upon the extensively adopted fractional calculus theory, we generalized the currently existing relationships such as Hansbo equation and Swartzendruber equation. The analytic solution for the fractional derivative flow model is acquired and the relative parameters are determined. Sets of experimental data are utilized to verify the validity of the proposed fractional derivative flow model. The comparative analysis demonstrates that including the Swartzendruber equation as a special case when the fractional derivative order $\gamma=1$, the fractional derivative flow model turns out to be a more flexible and accurate one to characterize the behavior of non-Darcian flow. Furthermore, a sensitivity study shows that the fractional derivative order is an essential parameter impacting the shape of $q-i$ curve. However, the corresponding physical interpretation of fractional derivative is not clear and further research is required to determine the relationship between the fractional derivative order and other mechanical parameters.
\section*{Acknowledgement}
The present work is supported by the National Natural Science Foundation of China (51674266), and the State Key Research Development Program of China (2016YFC0600704) and the Specialized Research Fund for the Doctoral Program of Higher Education (20130023110017). The financial supports are gratefully acknowledged. Special thanks are due to H.H. Liu, Aramco Services Company, for his valuable suggestions and help in improving the article.
\section*{References}
\bibliography{Bibfile}
\end{document}